\documentclass[aps,reprint,superscriptaddress,twocolumn,floatfix,showpacs,longbibliography]{revtex4-1}

\usepackage{graphicx,color}
\graphicspath{{images/}{figures/}{pictures/}}
\usepackage[utf8]{inputenc}
\usepackage{amsmath,amssymb,bm,amsfonts,dsfont,mathrsfs,amsthm}
\usepackage{braket}
\usepackage{txfonts,comment}
\usepackage[colorlinks=true,linkcolor=blue,citecolor=blue,urlcolor=blue]{hyperref}
\usepackage{soul}
\usepackage{enumitem}

\AtBeginDocument{%
\newwrite\bibnotes
\def\bibnotesext{Notes.bib}
\immediate\openout\bibnotes=\jobname\bibnotesext
\immediate\write\bibnotes{@CONTROL{REVTEX41Control}}
\immediate\write\bibnotes{@CONTROL{%
    apsrev41Control,author="08",editor="1",pages="1",title="0",year="1"}}
\if@filesw
\immediate\write\@auxout{\string\citation{apsrev41Control}}%
\fi
}%

\begin{document}
\title{Enhanced image classification via hybridizing quantum dynamics with classical neural networks}

\author{Ruiyang Zhou}
\email{ruiyangzhou@std.uestc.edu.cn}
\affiliation{Institute of Fundamental and Frontier Sciences, University of Electronic Science and Technology of China, Chengdu 611731, China}
\affiliation{Key Laboratory of Quantum Physics and Photonic Quantum Information, Ministry of Education, University of Electronic Science and Technology of China, Chengdu 611731, China}

\author{Saubhik Sarkar}
\email{saubhik.sarkar@uestc.edu.cn}
\affiliation{Institute of Fundamental and Frontier Sciences, University of Electronic Science and Technology of China, Chengdu 611731, China}
\affiliation{Key Laboratory of Quantum Physics and Photonic Quantum Information, Ministry of Education, University of Electronic Science and Technology of China, Chengdu 611731, China}

\author{Sougato Bose}
\email{s.bose@ucl.ac.uk}
\affiliation{Department of Physics and Astronomy, University College London, Gower Street, WC1E 6BT London, United Kingdom}

\author{Abolfazl Bayat}
\email{abolfazl.bayat@uestc.edu.cn}
\affiliation{Institute of Fundamental and Frontier Sciences, University of Electronic Science and Technology of China, Chengdu 611731, China}
\affiliation{Key Laboratory of Quantum Physics and Photonic Quantum Information, Ministry of Education, University of Electronic Science and Technology of China, Chengdu 611731, China}
\affiliation{Shimmer Center, Tianfu Jiangxi Laboratory, Chengdu 641419, China}

\begin{abstract}
The integration of quantum computing and machine learning has emerged as a promising frontier in computational science.
We present a hybrid protocol which combines classical neural networks with non-equilibrium dynamics of a quantum many-body system for image classification. 
This architecture leverages classical neural networks to efficiently process high-dimensional data and encode it effectively on a quantum many-body system, overcoming a challenging task towards scaled up quantum computation. 
The quantum module further capitalizes on the discriminative properties of many-body quantum dynamics to enhance classification accuracy. 
By mapping images from distinct classes to nearly-orthogonal quantum states, the system maximizes separability in the Hilbert space, enabling robust classification. 
We evaluate the performance of our model on several benchmark datasets with various number of features and classes. 
Moreover, we demonstrate the key role of the quantum module in achieving high classification accuracy which cannot be accomplished by the classical neural network alone. 
This showcases the potential of our hybrid protocol for achieving practical quantum advantage and paves the way for future advancements in quantum-enhanced computational techniques. 
\end{abstract}


\maketitle


\section{Introduction}
\label{sec:introduction}

The superiority of quantum technologies over classical counterparts has been demonstrated in several algorithms, including Shor's algorithm for integer factorization~\cite{shor1994algorithms}, Grover's algorithm for unstructured search~\cite{grover1996fast}, and the Harrow–Hassidim–Lloyd algorithm for solving linear systems~\cite{harrow2009quantum}. 
However, such algorithms with rigorously proven quantum advantage remain exceptionally rare, with only a handful known to date. 
Moreover, their large-scale realization demands fault-tolerant quantum computers~\cite{gottesman1997stabilizer, knill1997theory, kitaev2003fault}, whose practical realizations remain beyond the capabilities of near-term  quantum technologies. 
On the other hand, heuristic quantum algorithms - despite lacking rigorous proofs of quantum advantage - may represent a more practical way to ultimately outperforming classical algorithms~\cite{bharti2022noisy}. 
In particular, they are more friendly to be implemented on noisy intermediate scale quantum (NISQ) devices~\cite{preskill2018quantum}. 
Most prominent examples of heuristic algorithms are variational quantum algorithms (VQA)~\cite{wecker2015progress, mcclean2016theory,cerezo2021variational} which include: (i) variational quantum eigensolvers (VQE)~\cite{peruzzo2014variational, cerezo2022variational}  for solving condensed matter systems~\cite{Lyu2023symmetryenhanced,Han2024multilevel,BravoPrieto2020scalingof,Uvarov2020Variational,Okada2023Classically} as well as quantum chemistry problems~\cite{nam2020ground, Kandala2017Hardware, peruzzo2014variational, Arute2020Hartree};  (ii) quantum approximate optimization algorithm (QAOA)~\cite{farhi2014quantum} for combinatorial optimization problems such as quadratic binary unconstrained optimization (QUBO)~\cite{lucas2014ising} which can be applied to traveling salesman~\cite{kieu2019travelling} and max-cut problems~\cite{festa2002randomized}.
The performance of these algorithms depends on the quality of the NISQ hardware, which is expected to improve over next few years.
However, even with the assumption of perfect quantum devices, the existing algorithms can suffer from some fundamental problems such as costly measurements~\cite{flammia2011direct, haah2016sample}, difficulty in designing quantum circuits~\cite{Huang2022robust, Martyniuk2024QuantumArchitecture, Chivilikhin2007MoGVQE, Huang2024Adaptive}, and slow optimization due to `barren plateaus'~\cite{mcclean2018barren, cerezo2021cost,Larocca2025Barren}.
Therefore, it is timely to explore more heuristic algorithms and study the possibility of achieving quantum advantage.

One of the sectors that shows a lot of hope for the success of these heuristic algorithms is quantum machine learning, which has emerged as a promising frontier in technology by integrating quantum computing and machine learning (ML)~\cite{schuld2015introduction, biamonte2017quantum, wang2024comprehensive}.
Initially, classical ML techniques were being used for characterizing quantum systems such as phase transitions~\cite{carrasquilla2017machine, van2017learning}, designing multi-qubit gates~\cite{banchi2016quantum}, estimating entanglement~\cite{gray2018machine}, tomography~\cite{torlai2018neural, torlai2019integrating}, and ground state search~\cite{carleo2017solving, saito2017solving}, quantum sensing~\cite{Palmieri2021Multiclass,huang2025quantum,Gebhart2021Bayesian,nolan2021machine}, and even non-Hermitian systems~\cite{yu2021unsupervised, heredge2025nonunitary}.
However, quantum-inspired ML models~\cite{mangini2021quantum} were soon developed, which includes quantum perceptrons~\cite{schuld2015simulating}, quantum feed-forward neural network~\cite{mitarai2018quantum, perez2020data, schuld2021effect}, quantum convolutional neural network~\cite{grant2018hierarchical, cong2019quantum, henderson2020quanvolutional, pesah2021absence}, quantum recurrent neural network~\cite{bausch2020recurrent}, quantum reservoir learning~\cite{fujii2017harnessing, ghosh2019quantum, Palacios2024Role, Sannia2024dissipationas, Mujal2023TimeSeries, MartinexPena2021dynamical}, quantum extreme learning~\cite{mujal2021opportunities, zia2025quantumextreme, Suprano2024experimental, innocenti2023potential}, quantum ensemble learning~\cite{schuld2018quantum, Li2024Ensemble, Macaluso2024efficient}, dissipative quantum neural networks~\cite{beer2020training, sharma2022trainability}, kernel-based quantum algorithms~\cite{Huang2021Power, Daley2022PracticalQuantum, perezSalinaz2021one, Cerezo2023challenges, Jager2023Universal, Jager2023Universal} and quantum Boltzmann machines~\cite{amin2018quantum}. 
While the advantage of using quantum computers for processing quantum data (such as many-body ground states) is understandable~\cite{schuld2021supervised, wu2023quantum, khosrojerdi2025learning}, demonstrating a similar advantage for classical data via heuristic quantum algorithms remains an open challenge~\cite{Angrisani2024Classically,angrisani2025simulating,Martinez2025Efficient}. 
Indeed, achieving practical quantum advantage through heuristic algorithms relies on their implementation in large-scale quantum computers, which requires further improvement on NISQ devices. 
A key obstacle is the high dimensionality of classical data. 
Although quantum mechanics theoretically enables efficient encoding (e.g., through amplitude encoding), no practical realization exists to date.

Image processing and classification is one of the central tasks where the classical ML techniques are extensively used for practical purposes.
Therefore, it is highly desirable to potentially gain operational advantages originating from quantum processing.
Classification tasks have attempted with the aforementioned quantum-inspired neural networks, along with quantum principal component analysis (PCA)~\cite{lloyd2013quantum, huang2022quantum}, and quantum support vector machines (SVM)~\cite{rebentrost2014quantum, havlivcek2019supervised}.
However encoding the high-dimensional features and slowing down of optimization due to barren plateaus enforce us to continue looking for alternative strategies.

In this paper, we introduce a novel hybrid classical-quantum neural network architecture that integrates classical neural networks with time evolution of quantum systems.
Our approach performs the feature reduction of the high-dimensional image data with a simple classical neural network.
We then leverage non-equilibrium dynamics in many-body quantum systems for efficient classification.
This integration not only enables a reduction in the number of required qubits but also demonstrates superior accuracy and efficiency compared to classical counterparts. 
The remainder of this paper is organized as follows.
Section~\ref{sec:dynamics} describes the underlying Hamiltonian dynamics for discriminating the data.
Section~\ref{sec:protocol} details the architecture of our protocol.
Section~\ref{sec:data} describes the datasets used in our experiments.
Section~\ref{sec:simulations} presents our numerical evaluation on benchmark datasets including SAT6, BloodMNIST, MNIST, and Fashion-MNIST.
The factors analyzed include the measurement method, number of qubits, evolution time, number of samples for building observables, number of training samples, and number of hidden layers.
Section~\ref{sec:role} quantifies the role of the quantum dynamics in the classification task.
Finally, Section~\ref{sec:conclusion} concludes with discussions on the implications of our results as well as the broader implications and future research directions in quantum-enhanced image processing.

\section{Discriminative non-equilibrium dynamics}
\label{sec:dynamics}

\begin{figure}[t]
    \centering
    \includegraphics[width=0.85\linewidth]{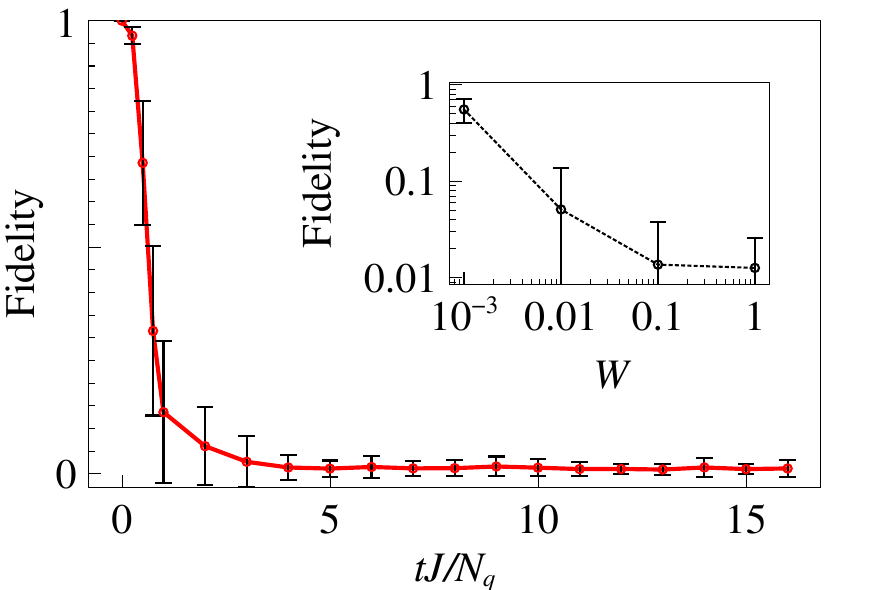}
    \caption{
        Evolution of fidelity between two states, starting identically, and evolving under the Hamiltonian in Eq.~\eqref{eq:hamiltonian} under two different sets of parameters.
        The average fidelity over 100 random simulations is shown as a function of rescaled time $tJ/N_q$ with number of qubits $N_q {=} 8$ and random magnetic filed strength $W {=} 1$.
        The inset shows the average fidelity at long time $tJ/N_q {=} 100$ as a function of $W$.
    }
    \label{fig:overlaps}
\end{figure}

Deterministic quantum state discrimination through performing a single measurement is only possible between orthogonal quantum states~\cite{barnett2009quantum}. 
Thus, by developing an encoding scheme that maps classical images to quantum states while ensuring almost orthogonality between states of different classes, one can efficiently use quantum systems for image processing. 
In this section, we show that non-equilibrium dynamics of quantum states is a powerful tool for achieving such encoding. 
Let's consider two quantum systems which both are initially in identical quantum states $\ket{\Psi_0}$ and therefore, one cannot discriminate them by any choice of measurements. 
However, if the two systems evolve with two slightly different Hamiltonians $H_1$ and $H_2$, i.e., $\ket{\Psi_j(t)}=e^{-iH_jt}\ket{\Psi_0}$ (for $j=1,2$), then their fidelity $F(t) {=} |\braket{\Psi_1(t) | \Psi_2(t)}|^2$ may decrease rapidly as a function of time by increasing the distance between $H_1$ and $H_2$. 
In other words, as time passes one can discriminate the quantum states $\ket{\Psi_1(t)}$ and  $\ket{\Psi_2(t)}$, which were initially indistinguishable, more efficiently. 
This means that by encoding classical images in the Hamiltonian, one can harness discriminative non-equilibrium dynamics of quantum systems. 
This can be illustrated through an example.  
We consider a chain of $N_q$ qubits evolving under the transverse field Ising Hamiltonian
\begin{equation}
    H = J \sum_{j=1}^{N_q-1} \sigma^x_{j} \sigma^x_{j+1} + \sum_{j=1}^{N_q} h_j \sigma^z_{j} \text{,}
    \label{eq:hamiltonian}
\end{equation}
where $J$ is the tunneling strength, $\sigma^{x,z}_{j}$ are the Pauli matrices at site $j$, and $h_j$ is the site-dependent magnetic field. 
The system is initially prepared in the quantum state $\ket{\Psi_0}{=}\ket{0,0,\cdots,0}$.
We look at the evolution of fidelity $F(t)$ between two states, which is initially $F(0) {=} 1$. 
The states are subsequently propagated with two sets of $\{h_j\}$ values such that each $h_j$ is drawn from a random uniform distribution $[-W, W]$ with strength $W$.    
In Fig.~\ref{fig:overlaps}, we plot $F(t)$ as a function of time, averaged over several random sets of $\mathbf{h}=(h_1,h_2,\cdots,h_{N_q})$. 
As the figure shows, the average fidelity exhibits a rapid drop from unity and goes towards zero as time increases, demonstrating that distinct sets of Hamiltonian parameters result in distinct states.
Small random field strength $W$ is sufficient to achieve this drop in fidelity, as shown in the inset. 
Therefore, by establishing a mapping between classical images and the set of magnetic fields $\mathbf{h}$, we can use this property for classification.

\section{Hybridizing quantum dynamics with classical neural networks for image processing}
\label{sec:protocol}

As explained in the precious section, non-equilibrium dynamics in a quantum system could be useful for creating orthogonal quantum states, which can be discriminated through quantum measurements. 
In order to exploit this for image classification, one has to establish a map between classical images, whose sizes are typically very large, into a set of Hamiltonian parameters, e.g. $\mathbf{h}{=}(h_1,h_2,\cdots,h_{N_q})$ in Eq.~(\ref{eq:hamiltonian}), which are typically very few. 
In this section, we put forward a protocol based on hybridizing classical neural networks with non-equilibrium dynamics of a simple transverse Ising Hamiltonian for achieving  good accuracy classification. 

\begin{figure*}[t]
    \centering
    \includegraphics[width=0.8\linewidth]{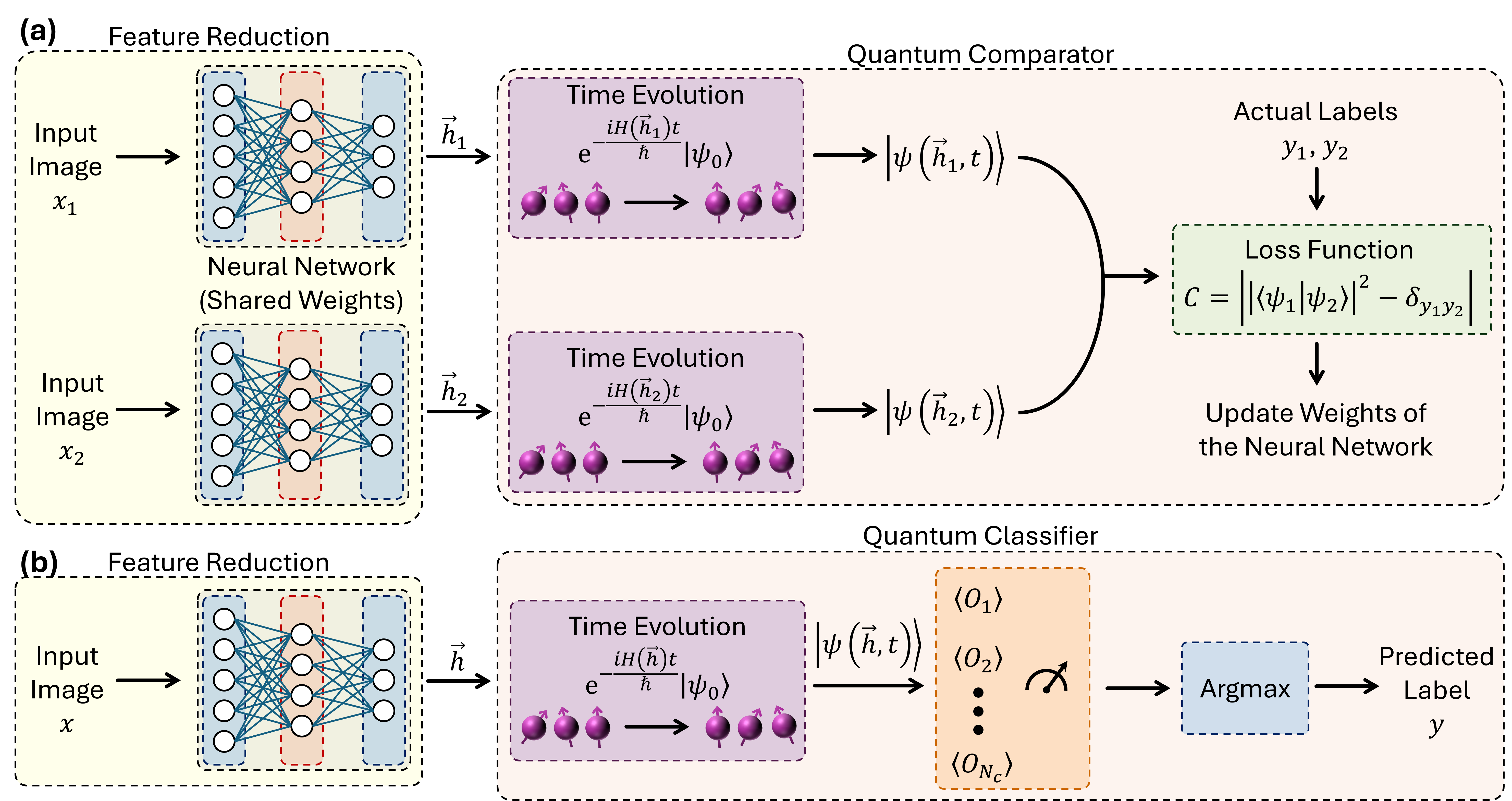}
    \caption{%
        Protocol by the hybrid classical-quantum neural network.
        (a) The quantum comparator.
        (b) The quantum classifier.
        Both the quantum comparator and quantum classifier start with a classical module, which is a lightweight classical neural network for feature extraction and dimensionality reduction.
        The classical module is followed by a quantum module, which is a quantum system that evolves the encoded data through Hamiltonian dynamics.
    }
    \label{fig:protocol}
\end{figure*}

\subsection{Modules of the protocol}

Our protocol is a hybrid structure of classical neural networks and quantum dynamics of an Ising transverse field system. 
This hybridization, combines the strengths of classical machine learning and quantum dynamics to efficiently capture complex patterns within image data. 
The protocol consists of two separate modules: (i) a classical neural network whose main task is feature reduction and is optimized during the training procedure; and (ii) an analog quantum system with transverse field Ising Hamiltonian in which the magnetic field strengths are determined by the output of the classical neural network.

We first discuss the classical module. 
Initially, each image is processed through a lightweight classical neural network, which extracts and compresses essential features from the input data.
This classical module serves as an encoder, reducing the dimensionality of the data and preparing it for quantum processing. 
The number of neurons in the input layer of the classical neural network is equal to the features (i.e., number of pixels) of the input image. 
The number of neurons in the output layer is equal to the number of qubits $N_q$ in the quantum Ising chain. 
For the sake of simplicity and the ease of training, we use only one hidden layer with 256 neurons between the input and output layers and consider the layers to be fully connected. 
The use of a fully connected neural network for dimensionality reduction offers several advantages over other techniques, such as center cropping, principal component analysis (PCA), or pooling operations like maximum pooling or average pooling.
The fully connected layers can learn an optimized transformation that captures the most relevant features from the input data, tailored to the specific task and dataset.
In addition, the non-linear activation functions in the fully connected layers allow the network to model complex,
non-linear relationships between the input and output dimensions, which may be beneficial for representing the intricate patterns present in the data.
Moreover, by incorporating the dimensionality reduction layers within the quantum neural network architecture, the parameters of these layers can be optimized jointly with the rest of the quantum system during the training process, potentially leading to better overall performance.

The second module of the protocol is an analog quantum simulator whose Hamiltonian is described by an Ising model with transverse field, as given in Eq.~(\ref{eq:hamiltonian}). 
While the exchange coupling $J{=}1$ is kept fixed, the magnetic fields of the Hamiltonian, namely the elements of the vector $\mathbf{h}$, are determined by the output of the classical module. 
The initial state of the qubits is always the ferromagnetic state $\ket{\Psi_0}$. 
The system evolves over a time $t$, harnessing the inherent quantum properties such as superposition and entanglement, which naturally arise during the evolution, to explore a richer feature space beyond the capacity of classical architectures.

\subsection{Training: a quantum comparator scheme}

In this section, we explain how we train the neural network as part of our hybrid structure.  
In the training procedure, our image classification protocol uses images from a training dataset with $N_c$ classes (or labels) and $N_f$ number of features (or pixels). 
The images are reshaped into $N_f$ dimensional vectors $\mathbf{x_j}$, while their corresponding labels $y_j$ are integer-valued, representing $N_c$ distinct classes. 
The training is pursued  according to the following steps:

\begin{itemize} [leftmargin=*]
    \itemsep0em
    \vspace{-\topsep}
    
    \item \textbf{Step 1:} The comparator processes the images in pairs.  
    For two images $\mathbf{x_1}$ and $\mathbf{x_2}$ with labels $y_1$ and $y_2$, a fully-connected classical neural network generates outputs $\mathbf{h_1}$ and $\mathbf{h_2}$, respectively. 
    The neural network has an input layer and an output layer with $N_f$ and $N_q$ neurons, respectively, and a hidden layer in between with 256 neurons. 
    
    \item \textbf{Step 2:} The $N_q$ dimensional vectors $\mathbf{h_1}$ and $\mathbf{h_2}$ serve as the site-dependent magnetic fields for the transverse Ising Hamiltonian in Eq.~\eqref{eq:hamiltonian}. 
    The quantum state representing each image is obtained by evolving an initial product state $\ket{\Psi_0} {=} \ket{0,0,\cdots ,0}$ as $\ket{\Psi_j(t)}{=}e^{-i H(\mathbf{h_j}) \, t} \ket{\Psi_0}$ for $j{=}1,2$.
    
    \item \textbf{Step 3:} The fidelity between the two final quantum states $F(t) {=} |\braket{\Psi_1(t) | \Psi_2(t)}|^2$ is computed through swap test~\cite{Buhrman2001Quantum} or shadow tomography~\cite{aaronson2018shadow, huang2020predicting} to quantify their similarity. 
    Then, the loss function is expressed as 
    \begin{align}
    \mathcal{C} {=} \frac{1}{N_{\rm trn}} \sum_{\{ \mathbf{x_1}, \mathbf{x_2} \}} \Big||\braket{\Psi_1(t) | \Psi_2(t)}|^2 - \delta{y_1, y_2} \Big| ,
    \end{align}
    where $N_{\rm trn}$ is the number of image pairs used for training.
    
    \item \textbf{Step 4:} Optimization is conducted by adjusting the  weights and biases in the classical neural network to ensure that fidelity accurately reflects the similarity or dissimilarity between image pairs. 
    We use stochastic gradient descent with adaptive momentum based method (Adam)~\cite{kingma2014adam} for the optimization.
\end{itemize}

In  Fig.~\ref{fig:protocol}(a), the schematic picture of the quantum comparator is depicted.

\subsection{Image  classification}

Once the training is over, we can use the hybrid setup for classifying an unseen data $\mathbf{\hat{x}_j}$ by assigning a label $\hat{y}_j$.  
Ideally the assigned label $\hat{y}_j$ has to be the same as the real label $y_j$. 
The procedure for assigning the label $\hat{y}_j$ is pursued through the following steps:

\begin{itemize} [leftmargin=*]
    \itemsep0em
    \vspace{-\topsep}
    
    \item \textbf{Step 1:} Building upon the comparator, the quantum classifier leverages previously trained classical module parameters. 
    A single test image $\mathbf{\hat{x}_j}$ is converted to a magnetic field vector by the trained classical neural network. 
    The corresponding Hamiltonian is then used for the dynamics of the system to create the corresponding quantum state $\ket{\hat{\Psi}_j(t)}$.
    
    \item \textbf{Step 2:} The final state $\ket{\hat{\Psi}_j(t)}$ is subjected to measurement of $N_c$ observables that are tailored to the classification tasks. 
    The observable for each class are constructed from $N_S$ number of projectors based on the final states obtained by taking $N_S$ samples from the training set for that class. 
    Explicitly, we write the observable for the class $\alpha$ as
    \begin{equation}
    \label{eq:observables}
    O_{\alpha} = \frac{1}{N_S} \sum_{k=1}^{N_S} \ket{\Psi_{k}^{(\alpha)} (t)}\bra{\Psi_{k}^{(\alpha)} (t)} ,
    \end{equation}
    where $\ket{\Psi_{k}^{(\alpha)} (t)}$ is the state obtained after evolution time $t$ with magnetic fields corresponding to the $k$-th sample image of  class $\alpha$ in the training set.
    Therefore, each observable acts as a characteristic representation of the quantum states associated with its corresponding class. 
    
    \item \textbf{Step 3:} Classification decision for assigning the label $\hat{y}_j$ is based on $N_c$ measurements $\langle O_{\alpha}\rangle{=}\langle \hat{\Psi}_j(t)|O_{\alpha}|\hat{\Psi}_j(t)\rangle $. 
    The assigned label $\hat{y}_j$ would be the same as the label of the class $\alpha$ whose corresponding averaged observable  $\langle O_{\alpha}\rangle$ has the largest value. 
\end{itemize}

By repeating the above procedure for a test set consisting of several unseen images, one can evaluate the performance of the classifier as
\begin{equation}
\label{eq:test_accuracy}
  \mathrm{Test\, \,  Accuracy}=\frac{1}{N_{\rm test}}\sum_{j=1}^{N_{\rm test}} \delta_{\hat{y}_j,y_j}
\end{equation}
where $N_{\rm test}$ is the total number of images in the test set. 
We will use this as a figure of merit for quantifying the quality of our classifiers.

\begin{figure*}[t]
    \centering
    \includegraphics[width=0.8\linewidth]{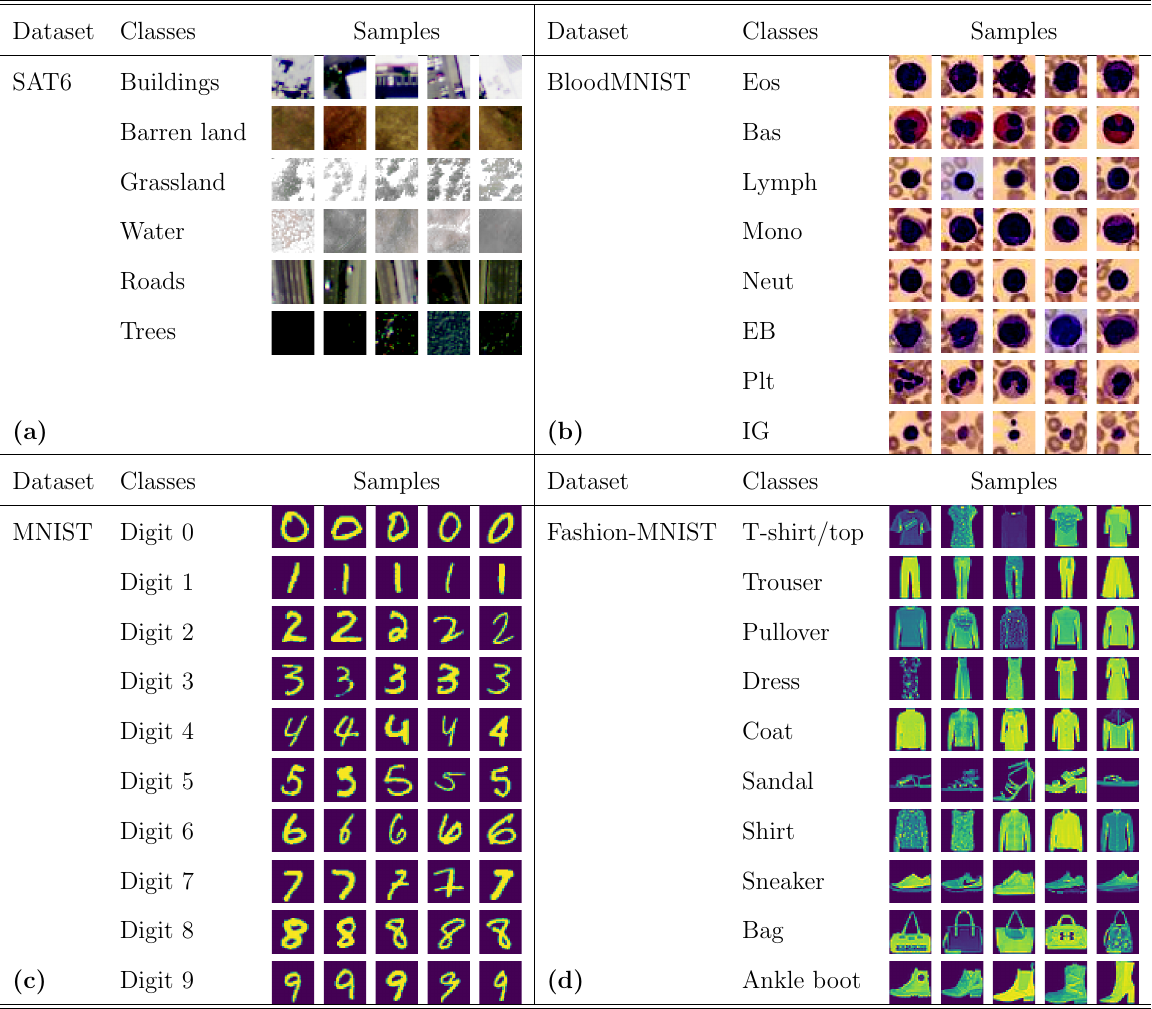}
    \caption{%
        Data samples from the (a) SAT6, (b) BloodMNIST, (c) MNIST, and (d) Fashion-MNIST datasets.
        (a) SAT6, containing six distinct land-cover classes (barren land, buildings, grassland, roads, trees, and water), each captured as a 28x28 pixels with 4-channel image.
        (b) BloodMNIST, comprised of eight classes of blood cell types (Eosinophil cells, Basophil cells, Lymphocyte cells, Monocyte cells, Neutrophil cells, Erythroblast cells, Platelet cells, and Immature Granulocytes cells),  depicted in 28$\times$28 RGB images specifically curated for diagnostic tasks.
        (c) MNIST, featuring 10 classes of handwritten digits (0--9) in 28$\times$28 pixel grayscale format, an extensively used benchmark in machine learning.
        (d) Fashion-MNIST, containing 10 classes of fashion items (T-shirt/top, trouser, pullover, dress, coat, sandal, shirt, sneaker, bag, and ankle boot), likewise presented as 28$\times$28 grayscale images.
        Each subfigure illustrates a small subset of the respective dataset's image samples, demonstrating the diversity and visual complexity of these classification tasks.
    }
    \label{fig:samples}
\end{figure*}

\subsection{Measurements and computation of fidelity }

Note that the observables $O_\alpha$, see Eq.~(\ref{eq:observables}), are in general highly non-local which means that measuring such operators might be challenging. 
One can instead divide such measurement into computing the fidelity between $\ket{\hat{\Psi}_j(t)}$, obtained from an unseen data, and a group of $N_s$ quantum states $\ket{\Psi_{k}^{(\alpha)} (t)}$ from each class $\alpha$. 
The computation of these fidelities can be done in the same way as the comparator scheme, namely either through swap test or efficient shadow tomography. 

There are two ways to get the fidelity between two quantum states $\ket{\psi_1}$ and $\ket{\psi_2}$ through swap test. 
The first approach, requires a controlled-swap gate~\cite{Buhrman2001Quantum} which is controlled by an ancilla qubit as 
$U_{\rm CS}{=}\ket{0}\bra{0}\otimes I_{1,2}{+}\ket{1}\bra{1}\otimes S_{1,2}$, where $I_{1,2}$ represents identity acting on the Hilbert space of the two systems and $S_{1,2}$ is the swap operator whose operation is described as $S_{1,2}\ket{\psi_1}\ket{\psi_2}{=}\ket{\psi_2}\ket{\psi_1}$. 
In order to compute the fidelity between $\ket{\psi_1}$ and $\ket{\psi_2}$ one can initialize the ancilla qubit in the state $\ket{+}{=}(\ket{0}{+}\ket{1})/\sqrt{2}$. 
Thus, the operation of the controlled-swap operator  on the entire system is given by
\begin{eqnarray}
    U_{\rm CS}\ket{+}\ket{\psi_1}\ket{\psi_2}&=& \frac{1}{\sqrt{2}}\ket{+}\left( \ket{\psi_1}\ket{\psi_2}+\ket{\psi_1}\ket{\psi_2}\right) \\ \nonumber
    &+& \frac{1}{\sqrt{2}}\ket{-}\left( \ket{\psi_1}\ket{\psi_2}-\ket{\psi_1}\ket{\psi_2}\right).
\end{eqnarray}
By performing a single qubit measurement on the ancilla qubit in $x$ basis, one gets the two probabilities as $p_{\pm}{=}\frac{1{\pm}|\langle \psi_1|\psi_2\rangle|^2}{2}$, from which one can get the fidelity between  the two quantum states.  

The second approach for performing the swap test avoids the ancilla qubit and the controlled swap gate operation.
Instead, it considers the swap operator $S_{1,2}$ as a Hermitian operator and thus a quantum mechanical observable. 
By measuring the average of the swap operator one gets $\bra{\psi_1}\bra{\psi_2}S_{1,2}\ket{\psi_1}\ket{\psi_2}{=}|\langle \psi_1|\psi_2\rangle|^2$ which directly gives the desired fidelity~\cite{banchi2016entanglement}. 
Measuring the swap operator between two many-body systems can be accomplished by a set of two-qubit measurements between the corresponding qubits in the singlet-triplet basis, which is an established technique in solid state~\cite{petta2005coherent, house2015radio, petersson2010charge, frey2012dipole, Colless2013Dispersive} and cold atoms~\cite{Rey2007Preparation, Trotzky2008time, Trotzky2010Controlling} systems.

Alternatively, one can compute the average of observables $\langle O_\alpha\rangle$ through shadow tomography~\cite{aaronson2018shadow, huang2020predicting, elben2022the}. 
A true reconstruction of quantum states requires full quantum-state tomography which relies on exponentially large number of measurements and thus is not scalable~\cite{Vogel1989Determination}. 
Shadow tomography offers a resource–efficient alternative to full quantum–state tomography when the goal is to predict the expectation values of many observables rather than to reconstruct the entire
density matrix~\cite{huang2020predicting}.
The protocol compresses information about an $N_q$‑qubit state $\rho$ into a collection of \emph{classical shadows} through three elementary steps:

\begin{enumerate}[leftmargin=*]
    \itemsep0em
    \vspace{-\topsep}
    
    \item \textbf{Step 1: Randomized measurements.}  
    For each measurement shot $m$ every qubit is acted on by a single‑qubit Pauli operator chosen randomly from $\{\sigma^{x,y,z}\}$ and is then measured in the computational ($\sigma^{z}$) basis. 
    The applied unitaries \(U^{(m)} {=} \bigotimes_{n}^{N_q} U^{(m)}_n\) and outcome binary strings $\mathbf{b^{(m)}} {=} (b^{(m)}_{1},\dots,b^{(m)}_{N_q})$ are recorded.
    
    \item \textbf{Step 2: Snapshot reconstruction.}  
    The average over these shots can be considered as a measurement channel.
    This channel can be inverted to reconstruct $\rho$ by mapping the Pauli operators in $U^{(m)}$ into new operators by replacing $\sigma^x$ by Hadamard,  $\sigma^y$ by phase-gate, and $\sigma^z$ by identity. 
    Thus, we obtain a new operator $U_{n}^{\prime(m)}$, namely $U_{n}^{(m)} \mapsto U_{n}^{\prime(m)}$.  
    As a result, each shot can be thought of as a state estimator
    \begin{equation}
       \rho^{(m)} = \bigotimes_{n=1}^{N_q} \Bigl(3\,U_{n}^{\prime\dagger(m)} \,\ket{b^{(m)}_{n}}\!\bra{b^{(m)}_{n}}\, U_{n}^{\prime(m)} -\mathds{1}\Bigr),
    \end{equation}
    whose ensemble average reproduces the true state, $\mathbb{E}[\rho^{(m)}]=\rho$ for enough number of measurements.

    \item \textbf{Step 3: Prediction of observables.}  
    With $M$ snapshots, the expectation value of an operator $O_{\alpha}$ is estimated as
    \begin{equation}
      \label{eq:shadow_pred}
      \langle O_{\alpha}\rangle = \frac{1}{M} \sum_{m=1}^{M} 
      \text{Tr} \Big(\rho^{(m)}O_{\alpha}\Big), \qquad j=1,\ldots,N_{c}.
    \end{equation}
\end{enumerate}

Note that a central merit of shadow tomography is its logarithmic sample complexity:
\(M=\mathcal{O}\!\bigl(\log N_{c}/\varepsilon^{2}\bigr)\) shots suffice to estimate \(N_{c}\) observables within additive error~$\varepsilon$, an exponential improvement over conventional tomography.  
This balance between accuracy and experimental overhead makes shadow tomography ideally suited for NISQ‑era devices and large‑scale simulations.
In this work we use $M {=} 5000$.

\section{Datasets}
\label{sec:data}

We now employ the classification protocol for four well-known datasets: SAT6, BloodMNIST, MNIST, and FashionMNIST.
Some sample images of each of these datasets are shown in Fig.~\ref{fig:samples}.
These datasets with different number of features are used to demonstrate the generalizability and broader application of the methodology in this study.

The SAT6 dataset~\cite{basu2015deepsat} in Fig.~\ref{fig:samples}(a) is a remote sensing image dataset composed of small images measuring $28\times28$ pixels with 4-channel (red, green, blue (RGB), and near infrared (NIR) for transparency), making the total number of features $N_f{=}3136$.  
With 324000 training images and 81000 testing images, the dataset contains $N_c{=}6$ distinct classes, each corresponding to different types of land cover.
This dataset is particularly useful for developing and benchmarking image classification algorithms in the field of remote sensing, where accurate identification of various land features is critical. 

The BloodMNIST dataset~\cite{yang2021medmnist, yang2023medmnist} in Fig.~\ref{fig:samples}(b) focuses on biomedical imaging and consists of images representing individual blood cells.
With 60,000 training images and 10,000 testing images, each image is $28\times28$ pixels in RGB channels  (i.e., the number of features $N_f{=}2352$)  belonging to one of the $N_c{=}8$ different classes.
This dataset is designed to support research in medical diagnostics, particularly for tasks involving the classification and analysis of blood cell types.

The MNIST dataset~\cite{xiao2017fashionmnist} in Fig.~\ref{fig:samples}(c) is a well-known collection of handwritten digits from $0$ to $9$ (i.e. the total number of classes is $N_c{=}10$) that has become a staple in the machine learning community.
It comprises 60,000 training images and 10,000 testing images, with each image being a $28\times28$ pixel, making the total number of features $N_f{=}784$. 
This dataset provides an ideal starting point for developing or testing new image recognition techniques, due to its simplicity and the extensive research already conducted using it.

The Fashion-MNIST dataset~\cite{li2012mnist} in Fig.~\ref{fig:samples}(d) offers another perspective on image classification, focusing on fashion items.
Like MNIST, it contains 60,000 training images and 10,000 testing images, with each image rendered in grayscale at $28\times28$ pixels (i.e., total number of features $N_f{=}784$) within $N_c{=}10$ different classes.
However, instead of digits, the images depict various fashion items, making it a popular alternative benchmark for testing models in computer vision.

\section{Numerical Results}
\label{sec:simulations}

\begin{table}[t]
    \begin{ruledtabular}
        \begin{tabular}{lccccc}
            Dataset       & $N_f$ & $N_c$ & $N_q$ & Test Acc. (\%) $\pm$ SEM \\
            \hline
            SAT6          & 3136  & 6     & 3     & 92.38 $\pm$ 1.33         \\
            BloodMNIST    & 2152  & 8     & 3     & 74.97 $\pm$ 0.93         \\
            MNIST         & 784   & 10    & 4     & 95.63 $\pm$ 0.21         \\
            Fashion-MNIST & 784   & 10    & 4     & 83.86 $\pm$ 0.44         \\
        \end{tabular}
    \end{ruledtabular}
    \caption{%
        Performance comparison of the protocol on the SAT6, BloodMNIST, MNIST, and Fashion-MNIST datasets with 
        $N_f$ features and $N_c$ number of classes.
        The accuracy is given as the mean test accuracy over 10 runs, with the standard error of the mean (SEM) reported.
        The model uses $N_q$ qubits, 50,000 training samples and $ 2 N_q $ evolution time.
    }
    \label{tab:acc2_baseline}
\end{table}

We evaluate the performance of the protocol on the SAT6, BloodMNIST, MNIST, and Fashion-MNIST datasets.
Table~\ref{tab:acc2_baseline} summarizes the classification accuracy achieved by the protocol on each dataset, along with the number of classes and qubits used in the quantum module.
The results demonstrate the effectiveness of the hybrid classical-quantum neural network architecture in accurately classifying images across a diverse range of datasets.

\subsection{Impact of the number of samples for measurement}

\begin{figure*}[t]
    \centering
    \includegraphics[width=0.75\linewidth]{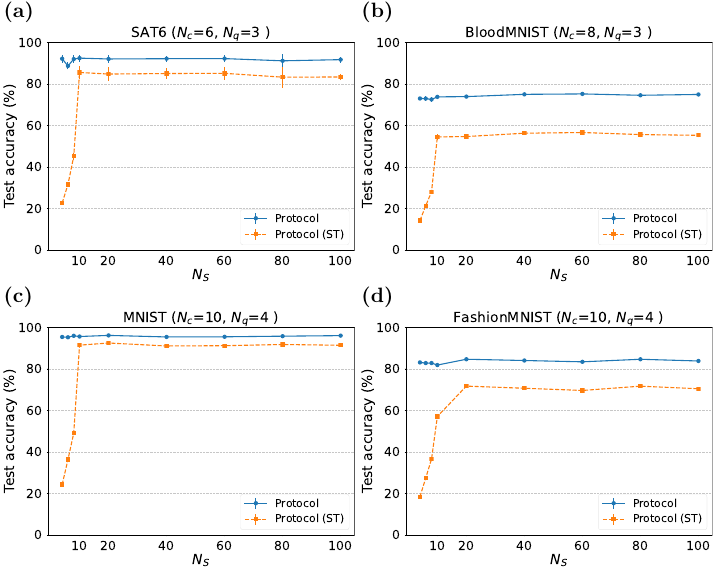}
    \caption{%
        Impact of the number of samples to construct the observables on the classification accuracy
        for the (a) SAT6, (b) BloodMNIST, (c) MNIST, and (d) Fashion-MNIST datasets.
        In all panels, blue circles (exact expectation) and orange squares (shadow tomography) represent the measurement methods used to obtain the classification accuracy.
        Here, $N_{\text{trn}} = 50,000, t = 2 N_q/J$.
    }
    \label{fig:acc2_test1}
\end{figure*}

One of the major resources that is used in our protocol is the number of measurement samples $N_S$ which is used for the construction of the observables $O_\alpha$ in Eq.~(\ref{eq:observables}). 
As discussed above, the averaged observables $\langle O_\alpha\rangle$ can be computed through either swap test or shadow tomography. 
In Figs~\ref{fig:acc2_test1}(a)-(d), we illustrate how the test accuracy of a quantum classifier depends on the number of final states $N_S$ used to construct each observable, under both swap test and shadow tomography for all the four datasets, respectively.
In practice, a smaller number of samples expedites classification task but may increase statistical noise, while a larger sample size yields more precise operator reconstructions at the cost of additional computational overhead. 
As the figure shows, using the swap test method remains largely unaffected by changes in sample size, consistently delivering high classification accuracy across all datasets. 
By contrast, shadow tomography requires sufficient sampling to converge. 
Three main observations can be made from Fig~\ref{fig:acc2_test1}. 
First, swap test is not only more efficient with respect to the number of measurement samples $N_S$ but also yields higher accuracy for any given number of measurement samples. 
Second, even in the shadow tomography approach, using a handful number of measurement samples (${\sim} 10$) is enough to converge to the best obtainable accuracy. 
This number for the swap test can be as small as $\sim 2$ measurement samples.   
Third, the overall behavior of the classifier also depends on the complexity of each dataset.
For relatively straightforward tasks such as MNIST, shadow tomography rapidly nears exact-measurement accuracy with a moderate increase in sample size. 
In more challenging domains like Fashion-MNIST, additional samples are needed to account for the dataset’s intricate visual features, resulting in a more gradual climb in performance.
BloodMNIST also shows a steady improvement with larger sample counts but retains a noticeable gap from the swap test approach, reflecting the subtleties of biomedical image classification.
These distinctions underscore how dataset complexity influences the sampling demands for tomography-based measurements.

\subsection{Impact of the evolution time}

\begin{figure*}[t]
    \centering
    \includegraphics[width=0.75\linewidth]{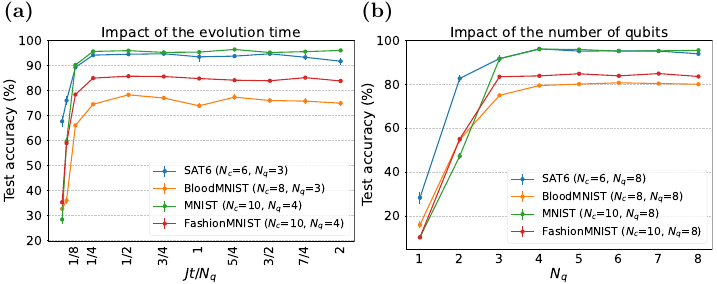}
    \caption{%
        The dependence of the classification accuracy for the SAT6, BloodMNIST, MNIST, and Fashion-MNIST datasets on (a) the evolution time and (b) the number of qubits.
        Error bars indicate statistical fluctuations across 10 runs.
        Here, $N_{\text{trn}} = 50,000, N_S = 100$.
    }
    \label{fig:acc2_time_qubits}
\end{figure*}

Another resource that is used in our system is the time that one needs to evolve the quantum system to achieve good accuracy. 
In Fig.~\ref{fig:acc2_time_qubits}(a) we illustrate how the test accuracy of a quantum classifier changes as a function of evolution time $t$ for four different datasets, with a fixed number of qubits and 50,000 training samples in each case.
We only show the results for the swap test approach in which we have taken $N_S {=} 100$, which is way beyond the requirement for the convergence. 
By measuring the classifier’s accuracy at each chosen $t$, we capture how longer or shorter evolution intervals affect the separability of different classes under both swap test and shadow tomography measurement schemes.
As the figure shows, for very short times, the accuracy is low as the non-equilibrium dynamics does not have enough time to achieve orthogonality between different quantum states. 
By increasing the time  greater than $\sim N_q/4J$, the accuracy becomes fairly stable and remains high. 
This is consistent with the results shown in Fig.~\ref{fig:overlaps} where quantum states become almost orthogonal after a similar time scale.
Across all cases, we observe that shadow tomography reliably captures the same effects of the evolution, while reaching close to the accuracy values achieved with exact measurement.
Therefore we only show the exact measurement results in Fig.~\ref{fig:acc2_time_qubits}(a) where we have taken $N_S {=} 100$ to make sure that both exact and shadow tomography calculations are independent of the sample size for constructing the observables

\subsection{Impact of the number of qubits}

The number of qubits in the quantum module is another important resource whose role has to be investigated. 
In a system with the size of $N_q$ qubits, one can find  $2^{N_q}$ orthogonal states, which is the size of the Hilbert space. 
Therefore, in order to get a good accuracy with our protocol one needs to satisfy the condition $2^{N_q}\ge N_c$ to allow that quantum states from different classes become orthogonal.  
In Fig.~\ref{fig:acc2_time_qubits}(b) we plot the the classification accuracy as a function of $N_q$ for the four selected datasets. 
As the figure shows, the obtainable accuracy for small system sizes that do not satisfy the above necessary condition is quite low. 
Only when the number of qubits $N_q$ exceeds $\log_2 N_c$ then the accuracy stabilizes. 
The rate of accuracy gain and the final performance levels differ among datasets, reflecting variations in data complexity and feature distribution.
In each case, the shadow-tomography approach closely matches the exact measurement, while the later is reported in Fig.~\ref{fig:acc2_time_qubits}(b) with $N_S {=} 100$.

\section{The role of quantum dynamics}
\label{sec:role}

\begin{figure}[b]
    \centering
    \includegraphics[width=0.8\linewidth]{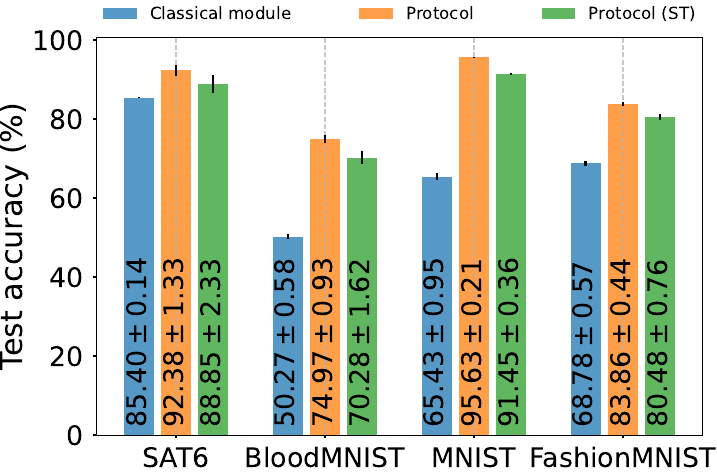}
    \caption{%
        The classification accuracy of only the classical module and the hybrid model for the SAT6, BloodMNIST, MNIST, and Fashion-MNIST datasets.
        The blue bar shows results with only the classical module, while the orange and green bars represent results with the hybrid model with the exact measurement and shadow tomography, respectively. 
        Here, $N_{\text{trn}} = 50,000, N_S = 100, tJ = 2 N_q$.
    }
    \label{fig:acc2_comparison}
\end{figure}

\begin{figure*}[t]
    \centering
    \includegraphics[width=0.95\linewidth]{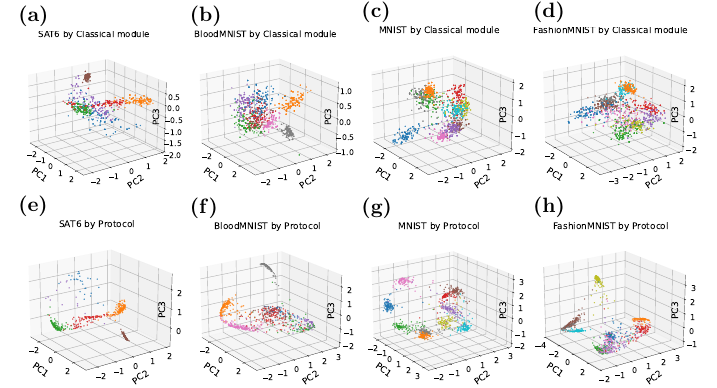}
    \caption{%
        Distinguishability comparison by the PCA on the SAT6, BloodMNIST, MNIST, and Fashion-MNIST datasets with $N_c$ number of classes.
        The model uses $N_q$ qubits, 50,000 training samples and $ 2 N_q $ evolution time.        
        $ N_{\text{trn}} = 50,000, N_S = 100, tJ = 2 N_q$.
    }
    \label{fig:pca3d}
\end{figure*}

Our protocol combines classical neural networks with non-equilibrium quantum dynamics of a transverse Ising spin chain. 
In this section we show that the role of the quantum module is crucial in enhancing the precision. 
The role of the non-equilibrium dynamics is separating the behavior of different $h$-fields was shown in Sec.~\ref{sec:protocol}A. 
Now we present the quantification of the merit of the quantum module in our protocol.

As a first check, we can solely use our classical neural network for solving the classification task. 
This shows the maximum capacity of the classical neural network that is used in our protocol for classifying the images. 
In order to have a fair comparison, we use the same number of neurons with the same geometry and only change the loss function to the standard mean squared error. 
Figure~\ref{fig:acc2_comparison} shows the comparison of the classification accuracy of only the classical module and the full hybrid model.
The crucial role of the quantum module in improving the performance of the hybrid model is clear from both the results obtained with the swap test as well as the shadow tomography. 
As the figure shows, the quantum module has different levels of enhancement for different datasets. 
This analysis clearly shows that the classical neural network which is used in this setup cannot reach the same accuracy of the quantum protocol even if we rely on shadow tomography to approximate our measured observables.

Further confirmation of the advantage of using the quantum module can be seen by performing principal component analysis (PCA) of the h-fields from the classical module and the measurement outputs of the full hybrid model.
PCA provides a statistical approach to capture the most important information from a high-dimensional data.
This is done by looking at the projection of each data along the eigenvectors with largest eigenvalues of the covariance matrix of the data.
As shown in Fig.~\ref{fig:pca3d}, while the outputs of the classical module are not quite distinguished by the PCA, the quantum module is able to classify the data in a lot more convincing manner.

\section{Conclusion}
\label{sec:conclusion}

Heuristic quantum algorithms are possibly the most viable approach towards achieving quantum advantage with NISQ devices. 
In this paper, we introduce a hybrid quantum-classical architecture for image classification problems which can handle large feature sets in the data, a major bottleneck that prevents quantum algorithms to be scaled up. 
By integrating classical neural networks with the discriminative power of non-equilibrium quantum many-body dynamics, we developed a protocol that effectively processes high-dimensional data and enhances classification accuracy through Hilbert-space separability. 
The system encodes images from different classes into nearly-orthogonal quantum states, enabling effective discrimination between them. 
We evaluated the performance of the protocol over four different datasets with various number of features and classes. 
Furthermore, we showed that the quantum module plays a key role in enhancing the classification accuracy beyond the capabilities of the classical module alone.  
These results not only demonstrate the viability of quantum-enhanced learning in the NISQ era but also pave the way for further exploration of quantum many-body systems in practical computational applications.

\begin{acknowledgments}

SS acknowledges support from National Natural Science Foundation of China (Grant No.~W2433012).
AB acknowledges support from National Natural Science Foundation of China (Grants Nos.~12050410253, 92065115, and 12274059). SB acknowledges the UKRI EPSRC grants Nonergodic quantum manipulation EP/R029075/1 and  Many-Body Phases In Continuous-Time Quantum Computation EP/Y004590/1 for support.

\end{acknowledgments}

\bibliography{Ref}

\newpage

\renewcommand{\thesection}{\Alph{section}}
\renewcommand{\thefigure}{S\arabic{figure}}
\renewcommand{\thetable}{S\Roman{table}}
\setcounter{figure}{0}
\renewcommand{\theequation}{S\arabic{equation}}
\setcounter{equation}{0}

\onecolumngrid
\section*{Supplementary Information}
\twocolumngrid
\subsection{Data pre-processing}

For neural networks, the data pre-processing steps involve converting the raw data into a suitable format for efficient processing.
In this study, we work with image datasets, and the following steps are performed:

\begin{enumerate} [leftmargin=*]
    \itemsep0em
    \vspace{-\topsep}

    \item \textbf{Data normalization:}
    This preprocessing step ensures that the input features are on a similar scale, which can improve the convergence rate and generalization performance of the models.
    In this study, we employ standard PyTorch normalization to shift and scale the image data to be in the range $[-1, 1]$. 
    
    \item \textbf{Conversion to Tensors}: The raw image data is converted into PyTorch tensors, which are multidimensional arrays that facilitate efficient computations on GPUs or other hardware accelerators.
    
    \item \textbf{Dataset Creation}: The tensor data is organized into a PyTorch Dataset, that provides a convenient way to access and iterate over the data samples during training and evaluation.
    
    \item \textbf{DataLoader Creation}: PyTorch DataLoaders are created from the Datasets, enabling efficient batching, shuffling, and parallel loading of data during the training process.
\end{enumerate}

\subsection{Impact of the number of hidden layers}

\begin{figure}[b]
    \centering
    \includegraphics[width=0.8\linewidth]{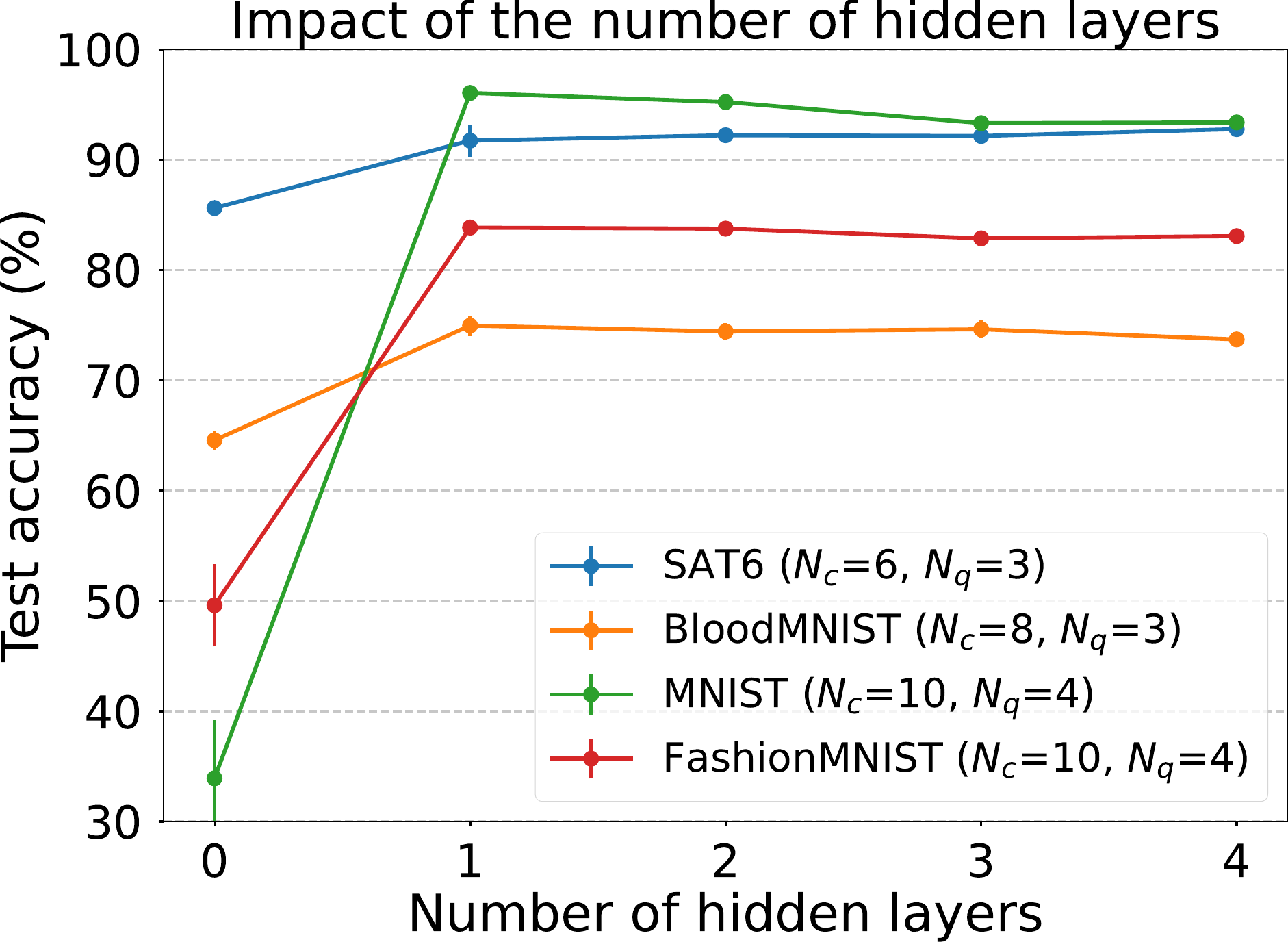}
    \caption{%
        Impact of the number of hidden layers on the classification accuracy for the four datasets.          
        Here, $ N_{\text{trn}} {=} 50,000, N_S {=} 100, tJ {=} 2 N_q$.
    }
    \label{fig:acc2_hidden}
\end{figure}
To investigate the impact of network depth on classification performance, we systematically varied the number of hidden layers in the classical neural network from zero up to four as shown in Fig.~\ref{fig:acc2_hidden}.
Each additional hidden layer introduces more parameters and non-linear transformations, enabling the model to learn increasingly complex features.
This procedure is straightforward: the architecture begins with a fixed input layer (mapped from the quantum measurement outcomes) and a fixed output layer (providing class probabilities), while the hidden layers in between are incrementally added or removed to examine how deeper or shallower designs influence overall accuracy.

Overall, introducing even a single hidden layer markedly boosts the test accuracy for both exact measurement and shadow-tomography approaches.
As more layers are added, performance tends to increase and eventually saturates, suggesting that beyond a certain depth, the benefit of additional layers diminishes.
These observations confirm that a sufficiently deep classical neural network is crucial for learning robust representations, but also indicate that excessively deep models may offer diminishing returns once the model capacity surpasses the complexity of the underlying data.
In our final protocol used in this paper, we have one hidden layer containing 256 neurons.

Despite these shared trends, the degree of improvement varies among datasets, largely due to differences in data complexity and feature diversity.
For instance, SAT6 and MNIST show rapid accuracy gains with just one or two hidden layers, indicating that relatively shallow architectures can already capture the most discriminative features.
BloodMNIST, by contrast, benefits more from deeper models, presumably because biomedical images often contain subtler distinctions that require more nuanced feature extraction.
Similarly, while Fashion-MNIST also sees steady improvements with increasing depth, it retains a gap between exact and approximate measurements, illustrating that some datasets demand higher fidelity to resolve fine-grained visual patterns.

\subsection{Impact of the number of training samples}

\begin{figure}[b]
    \centering
    \includegraphics[width=0.8\linewidth]{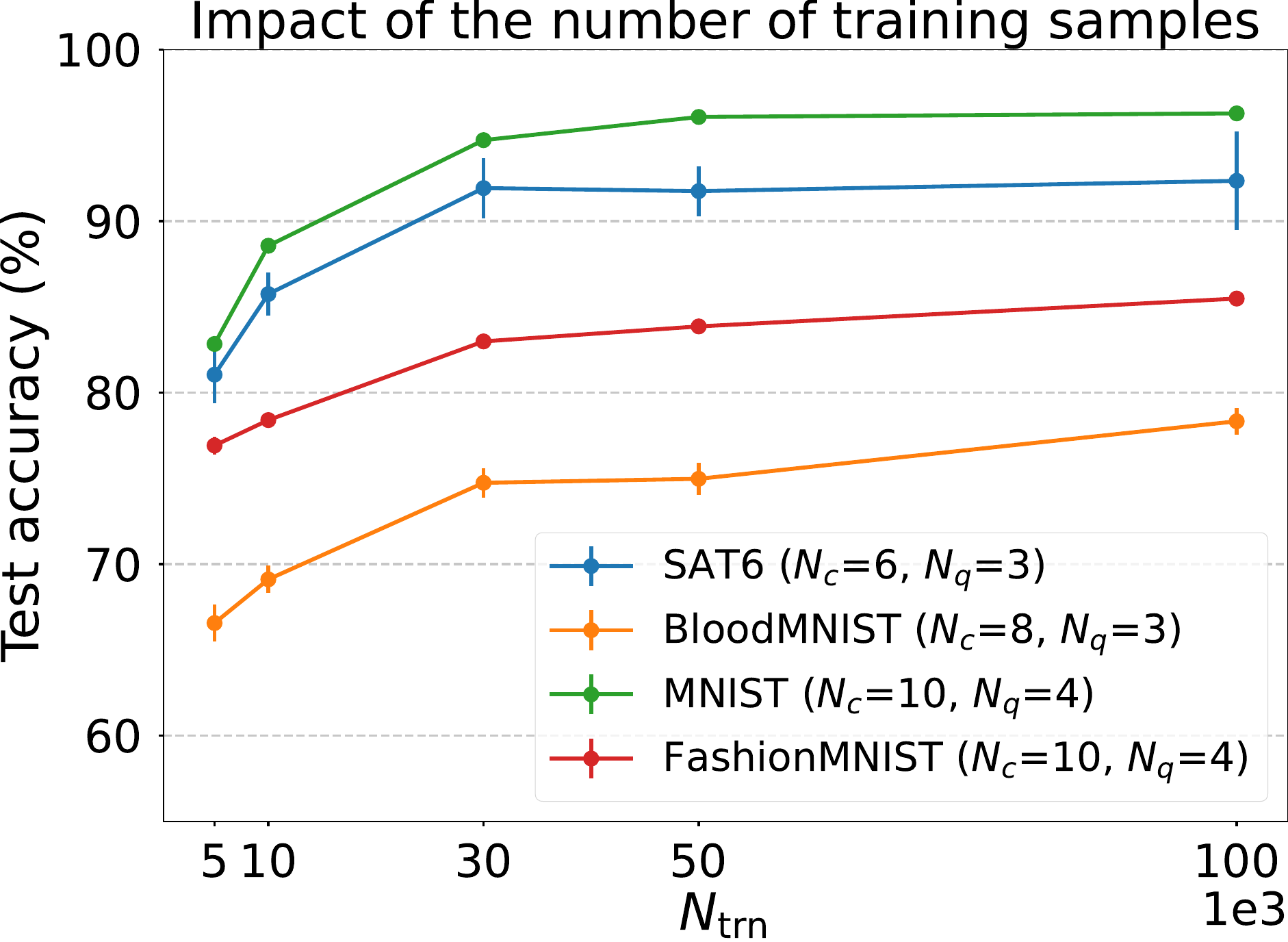}
    \caption{%
        Impact of the number of training samples on the classification accuracy for the four datasets.
        Here, $ N_S = 100, tJ = 2 N_q$.
    }
    \label{fig:acc2_train}
\end{figure}

Figure~\ref{fig:acc2_train} depicts how the test accuracy of a quantum classifier changes when the size of the training set is varied, while the number of qubits is kept fixed for each dataset.
Larger training sets consistently produced higher accuracy across all datasets and measurement methods.
Exact measurement invariably outperformed shadow tomography, but the gap between these two approaches often shrank as more samples were introduced. 
This trend highlights a fundamental principle in machine learning: increasing the volume of labeled data enriches the model’s representation of the input space, thereby strengthening its classification boundaries and enhancing generalization to unseen examples.

Nonetheless, the degree and pace of improvement varied among the four datasets.
For instance, SAT6 quickly reached near-saturation accuracy with relatively few samples, reflecting the dataset’s comparatively straightforward class distinctions.
By contrast, BloodMNIST and Fashion-MNIST exhibited a more gradual climb in performance and maintained a larger accuracy gap between exact and approximate measurements, suggesting that these tasks may require both more data and higher-fidelity measurements to capture their more nuanced feature distributions.
Meanwhile, MNIST’s inherently simpler digit shapes allowed the classifier to achieve high accuracy even with minimal training data, underscoring the role of dataset complexity in shaping learning outcomes.

\subsection{Principal Component Analysis}

The principal component analysis (PCA) can be used to extract the most informative features of a set of data vectors.
The first step of PCA is to normalize the data: $\mathbf X \rightarrow \tilde{\mathbf X}$.
One then computes the covariance matrix: $\mathbf S=\frac{1}{n-1}\tilde{\mathbf X}^{\!\top}\tilde{\mathbf X}$.
Eigen‑decomposition of the $d$-dimensional matrix gives the eigen-vector components in the descending order of the eigen-values: $\mathbf S\,\mathbf v_j=\lambda_j\mathbf v_j$ with $\lambda_1\ge\cdots\ge\lambda_d$
The first \(k\) vectors are chosen as the most important or principal components: $ \mathbf W_k=\bigl[\mathbf v_1\,\dots\,\mathbf v_k\bigr]$.
Now the data vectors are transformed by keeping only the first $k$ eigenvectors: $\mathbf Z_k=\tilde{\mathbf X}\,\mathbf W_k$, with variance along the components given by the corresponding eigenvalues. 

\end{document}